\documentclass[aps,prl,preprint,superscriptaddress]{revtex4}

\usepackage{epsfig} 

\begin{document}

\title{Critical Current of Type-II Superconductors in a Broken Bose Glass State}

\author{J. P. Rodriguez}
\affiliation{Department of Physics and Astronomy, 
California State University, Los Angeles, California 90032}

\date{\today}

\begin{abstract}
The tilt modulus of  a defective Abrikosov vortex lattice 
pinned by material line defects is computed
using the boson analogy.  
It tends to infinity at long wavelength,
which yields a Bose glass state 
that is robust to the addition of weak point-pinning centers,
and which implies a restoring force per vortex line
for rigid translations about mechanical equilibrium
that is independent of magnetic field.
It also indicates that the Bose glass state breaks into pieces along
the direction of the correlated pinning centers if the latter have finite
length.  The critical current is predicted to crossover from
two dimensional to three dimensional behavior
as a function of sample thickness along the correlated pinning centers
in such case.  That crossover notably can occur at a film thickness that is much
larger than that expected from point pins of comparable strength.
The above is compared to the dependence on thickness shown by the
critical current in certain films of high-temperature superconductors
currently being developed for wire technology.
\end{abstract}

\maketitle

\section{Introduction}
It well known that thin films of high-temperature superconductors exhibit
larger critical currents than their single-crystal counterparts.
Thin films of the high-temperature superconductor
YBa$_2$Cu$_3$O$_{7-\delta}$ (YBCO)
grown by pulsed laser deposition (PLD), 
which are actively being developed for wire technology,
%in particular,
achieve critical currents that are a significant fraction
of the maximum depairing current, for example\cite{dam}. 
Evidence exists that lines of dislocations that run parallel to the crystalline $c$ axis in PLD-YBCO
act as correlated pinning centers for vortex lines inside of the superconducting state\cite{klaassen},
%hence 
and thereby give rise to such high critical currents.
This is confirmed by  
the peak observed in the critical current of PLD-YBCO at orientations of
the $c$ axis aligned parallel to an applied magnetic field,
as well as by the dramatic enhancement of the former peak 
after more  material defects
in the form of nano-rods
aligned  parallel to the $c$ axis
are added\cite{bzo}\cite{goyal}. 
%Last, the introduction  of  additional  material line defects along the $c$ axis of PLD-YBCO  
%dramatically enhance the latter peak in the critical current versus film orientation\cite{bzo}\cite{goyal}.

The microstructure described above for PLD-YBCO films
immediately suggests that the vortex lattice that emerges
from the superconducting state
in applied magnetic field aligned parallel to the $c$ axis  
is some form of Bose glass  characterized by a divergent tilt modulus\cite{N-V}.
In the limit of a rigid vortex lattice, to be expected at large magnetic fields,
two-dimensional (2D) collective pinning of vortex lines
by the material line defects then determines the critical current
\cite{LO}\cite{kes-tsuei}\cite{M-E}\cite{Tink}.
Recent theoretical calculations that follow this line of reasoning
find moderate quantitative agreement with the critical current measured in films of PLD-YBCO
in $c$-axis magnetic fields  of a few to many kG, at liquid nitrogen temperature\cite{jpr-maley}.
A potential problem with the Bose glass hypothesis, however, is that the material line defects in
PLD-YBCO films can meander, or they can be of relatively short length\cite{bzo}\cite{goyal}.
That question is addressed in this paper,
where we find that the Bose glass {\it breaks up} into pieces
along the direction of the correlated pinning centers
when the effective length of the latter is less than the film thickness.  
In particular, the profile of the critical current versus film thickness
is predicted to reflect a  crossover from
two-dimensional to three-dimensional (3D) collective pinning of the vortex lines\cite{wordenweber-kes}.
This cross-over can occur at a length scale that is notably much larger than that
expected from point pins of comparable strength\cite{kim-larbalestier}.
We also find that the unbroken Bose glass state is robust to the addition
of weak point pins.
Good     agreement is achieved between the dependence on thickness 
shown by the critical current in certain films of PLD-YBCO 
at applied magnetic field\cite{kim-larbalestier} 
and that predicted for the true Bose glass state\cite{jpr-maley}.
Last, the effective restoring force per vortex line
due to a rigid translation of the Bose glass about mechanical equilibrium,
which is  gauged by the Labusch parameter\cite{koopmann},
is found to depend only weakly on applied magnetic field.  This prediction agrees with recent
measurements of the microwave surface resistance on PLD-YBCO films with nano-rod inclusions\cite{pompeo}.

\section {Tilt modulus of Bose glass}
\begin{figure}
\includegraphics[scale=0.8, angle=-90]{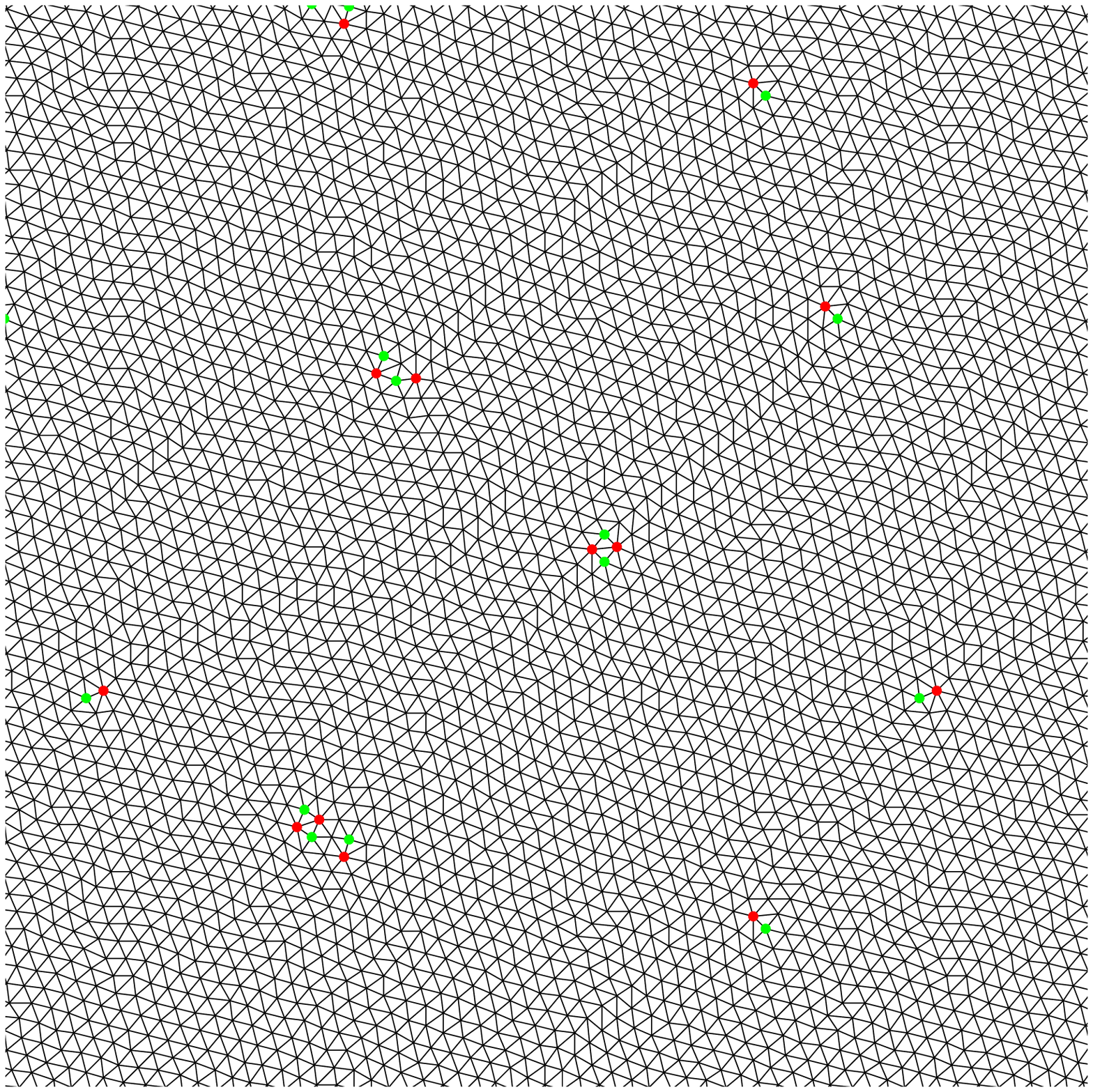}
%\includegraphics[scale=0.8, angle=-90]{xyt1u130}
%\centerline{\epsfxsize=60mm \epsfbox{FG42.ps}}
\caption{Shown is a Delaunay triangulation
of a   low-temperature groundstate
made up  of 2016 vortices 
that interact logarithmically over a 336 $\times$ 336 grid with periodic boundary conditions,
and that experience an equal number of identically weak $\delta$-function pinning centers
arranged in  a ``liquid'' fashion. (See ref. \cite{cec-jpr03}.)
The state shows macroscopic phase coherence and long-range hexatic order, 
with a supefluid density and a hexatic order parameter, respectively,
that are  29\% and 58\%
of the maximum possible values
attained by the perfect triangular vortex lattice.
It was obtained after simulated annealing from the liquid state
down to low temperature,
which resulted in 371 pinned vortices.
A red and green pair of disclinations forms a  dislocation (see ref. \cite{book}).
The above Monte Carlo simulation results are consistent with theoretical predictions
of a hexatic vortex glass state in two dimensions, in the  zero-temperature limit,
for pinning arrangements that do not show
any clusters or voids (ref. \cite{jpr05}).
Josephson coupling between layers then produces a Bose glass transition above zero temperature
as long as the 2D glass transtion is second order (ref. \cite{jpr04c}).}
\label{hex-glass}
\end{figure}

Material line defects in thin enough films of PLD-YBCO can be considered 
to be perfectly parallel to the $c$-axis.
They notably arrange themselves in a manner that
resembles a snapshot of a 2D liquid\cite{klaassen},
as opposed to a gas\cite{menghini}\cite{dasgupta}.
In particular, such correlated pinning centers do not show clusters or voids.
A defective  vortex lattice that  assumes a  
{\it hexatic} Bose glass state will then occur
for external magnetic fields aligned in parallel to 
such a microstructure, in the limit of weak correlated  pinning.
It is characterized by parallel lines of edge dislocation defects
that are injected into the pristine vortex lattice in order
to relieve shear stress due to the correlated pins. 
The former do {\it not} show any intrinsic tendency to arrange themselves into grain boundaries,
however, due to the absence of clusters and voids in the ``liquid''  arrangement
of linear pinning centers.
(Cf. refs. \cite{menghini} and \cite{dasgupta}.)
This results in a vortex lattice whose translational order is
destroyed at long range 
by the isolated lines of edge dislocations,
but which retains long-range orientational order.
Collective pinning
of the dislocation defects
by the correlated pins
then results both in an elastic response to shear 
and  in a net superfluid density.
%\cite{jpr05}.
% near zero temperature
Theoretical calculations\cite{jpr05} and
numerical Monte Carlo simulations (see fig. \ref{hex-glass} and caption)
of the corresponding 2D Coulomb gas ensemble\cite{cec-jpr03} 
confirm this picture.

The correlation length $L_c(|)$ for order along the magnetic field direction is infinite
for a Bose glass state.
In the limit of weak correlated pins,
the density of dislocation defects  that thread the corresponding  vortex lattice
% in the hexatic Bose glass state
can then be obtained by applying the 
theory of 2D collective pinning\cite{LO}\cite{kes-tsuei}\cite{M-E}.
Each line of unbound edge dislocations is in one-to-one correspondence with
a well-ordered bundle of vortex lattice, or Larkin domain, 
that has dimensions $R_c(|) \times R_c(|)$ 
in the directions transverse to the magnetic field.
The injection of the dislocation lines into 
the pristine vortex lattice then results in plastic creep
of each Larkin domain by a Burgers vector\cite{book}. 
The transverse Larkin scale is hence obtained by minimizing the sum of the elastic energy cost
due to the edge dislocations with the gain
in pinning energy due to the translation of a Larkin domain
by an elementary Burgers vector 
of the triangular vortex lattice,
$b = a_{\triangle}$.
%which is achieved by injecting lines of dislocations into the vortex lattice. 
This yields the estimate\cite{LO}\cite{kes-tsuei}\cite{M-E}
\begin{equation}
R_c(|)^{-2}   =
  C_0^2 n_{\rm p} (f_{\rm p} / c_{66} b)^2,
\label{ratio}
\end{equation}
for the density of Larkin domains,
which coincides with the density of lines of unbound dislocations.
Here $n_{\rm p}$ denotes
the density of pinned vortex lines,
$f_{\rm p}$ denotes the
maximum pinning force along a  material line defect per unit length,
and 
$c_{66}$ denotes the elastic shear modulus of the pristine vortex lattice.
The prefactor above is of order\cite{M-E}
$C_0 \cong  \pi /   {\rm ln} (R_c / a_{\rm df}^{\prime})^2$,
where  $a_{\rm df}^{\prime}$ is the
core diameter of a dislocation in the vortex lattice.
Consider now
the limit of weak pinning centers
that do not crowd together:
$f_{\rm p}\rightarrow 0$ and $\pi r_{\rm p}^2\cdot n_{\phi} \ll 1$, respectively,
where $n_{\phi}$ denotes the density of material line defects,
and where $r_{\rm p}$ denotes their range.
% of such  pinning centers.
Simple considerations of probability then yield the
identity $n_{\rm p}/ n_{\phi} = n_{\rm B} \cdot \pi r_{\rm p}^2$
between the fraction of occupied pinning centers and the product of
the density of vortex lines, 
$n_{\rm B}$, with
the effective area of each pinning center.
Substituting it plus the estimate 
$c_{66} = (\Phi_0 / 8 \pi \lambda_L)^2 n_{\rm B}$
for the shear modulus of the pristine vortex lattice\cite{brandt77}
into Eq. (\ref{ratio}) then yields the result
$R_c(|)^{-2} 
\cong (\sqrt{3}\pi/2) C_0^2
  (4f_{\rm p} r_{\rm p}/\varepsilon_0)^2 n_{\phi}$
for the density of Larkin domains,
which depends  only weakly on magnetic field.
Here, 
$\lambda_L$ denotes the London penetration depth and
$\varepsilon_0 = (\Phi_0/4\pi \lambda_L)^2$ 
is the maximum tension of  a flux line in the superconductor.
All of the above is valid
in the 2D collective pinning regime that exists
at perpendicular magnetic fields beyond the threshold
$B_{\rm cp} = C_0^2
(\sqrt{3} / 2)
(4 f _{\rm p} / \varepsilon_{0})^2 
\Phi_0$,
in which case many vortex lines are pinned by material line defects
within  a Larkin domain of transverse dimensions $R_c(|) \times R_c(|)$ \cite{jpr-maley}.

We will now exploit the boson analogy for vortex matter
% discovered by Nelson
in order to compute the uniform tilt modulus of the hexatic Bose glass in the
absence of point pinning centers\cite{N-V}.  
It amounts to a London model set by the free-energy density
\begin{equation}
g_{\rm B}(\{{\bf r}\},z) = \sum_i {1\over 2}\tilde\varepsilon_l\Biggl({d{\bf r}_i\over{d z}}\Biggr)^2
%+ \sum_{i > j} K_0 (|{\bf r}_i  - {\bf r}_j|/\lambda_L)
+ \sum_{i > j} V_0 ({\bf r}_i , {\bf r}_j)
+ \sum_i V_{\rm p}({\bf r}_i) 
\label{energy}
\end{equation}
for vortex lines located at transverse positions $\{ {\bf r}_i (z) \}$,
at a coordinate $z$ along the field direction.  
Here $\tilde\varepsilon_l$ denotes the tension of an
isolated vortex line, while the pair potential $V_0({\bf r}, {\bf r}^{\prime})$
describes the interaction between vortex lines at the same longitudinal coordinate $z$.
% over a scale $\lambda_L$ equal to the London penetration depth.  
The energy landscape for the correlated pinning centers is described by the
potential energy $V_{\rm p} ({\bf r})$,
which is independent of the coordinate $z$ along the field direction.
%Finally, 
The thermodynamics of this system in the presence of an external tilt stress
$n_{\rm B} {\bf a}$ is then set by the partition function
\begin{equation}
Z_{\rm B}[{\bf a}] = (\Pi_i\int{\cal D} [{\bf r}_i(z)]){\rm exp}
[-(k_B T)^{-1}\int_0^{L_z} dz [g_{\rm B} (\{{\bf r}\}, z) - \int d^2 r\, {\bf j}_{\rm B}\cdot {\bf a}]] ,
\label{z_b}
\end{equation}
which under periodic boundary conditions,
${\bf r}_i (z + L_z) = {\bf r}_i (z)$,
is equivalent to a system of 2D bosons.
Above, 
% $n_{\rm B}$ denotes the density of vortex lines, while
${\bf j}_{\rm B} ({\bf r}, z) = \sum_i \delta^{(2)}[{\bf r} - {\bf r}_i(z)] (d {\bf r}_i / d z)$
is the current density within the boson analogy.
Observe now that the kernel $\Pi_{\mu,\nu}(\omega)$ of
the uniform  electromagnetic response for alternating current (AC)
% mode
%$\Pi_{i,j}({\bf k}, i \omega_n)$ 
%(within the boson analogy) 
is connected to this partition function 
through the proportionality relationship
\begin{equation} 
Z_{\rm B} [{\bf a}] \propto 
{\rm exp} \Bigl[(k_{\rm B} T)^{-1} V\sum_{i\omega_n} 
{1\over 2} {\bf a}\cdot{\bf\Pi}\cdot{\bf a} \Bigr]
\label{pi}
\end{equation}
in the limit that the corresponding uniform tilt stress  vanishes,
${\bf a}\rightarrow 0$.  
Above,
the Matsubara frequencies
% ${\bf k}$ and 
$i \omega_n$
are given by the  allowed wavenumbers 
% perpendicular and 
along the magnetic field,
$q_z = 2\pi n/L_z$,
%, respectively.
and $V = L_x L_y L_z$ is the volume of the system.
The fact that the tilt stress is given by $n_{\rm B} {\bf a}$
then yields the identity
\begin{equation}
C_{44}(q_z) = n_{\rm B}^{2} / \Pi_{\perp} (\omega_n)
\label{identity}
\end{equation}
between the uniform tilt modulus and the uniform AC electromagnetic response of the 2D Bose glass.
%, respectively.
The subscript ``$\perp$'' above is a tag for the pure shear component of the electromagnetic kernel
$\Pi_{\mu,\nu}$ (Cf. ref. \cite{N-V}).

The hexatic Bose glass state is clearly a 2D dielectric insulator within the boson analogy.
Its electromagnetic response can therefore be modeled by the kernel
\begin{equation}
\Pi_{\perp} (\omega) = (n_{\rm B}/\tilde\varepsilon_l) \omega^2/(\omega^2 - \omega_0^2),
\label{pi}
\end{equation}
%
%Within the boson analogy,
which is dielectric in the low-frequency limit,  
and which conserves charge by satisfying  the oscillator f-sum rule.
Above, $\omega_0$ is the natural frequency
of the electric dipole degrees of freedom in the Bose glass.
The latter correspond to the 2D Larkin domains in reality,
since they represent 
%(plastically) mobile 
the smallest units of well-ordered vortex lattice
that can respond independently to an applied force.  
We then have that the above natural frequency
is of order the resonant frequency for transverse sound 
inside a Larkin domain of the 2D Bose glass:
$\omega_0 \sim \gamma / R_c (|)$, where
$\gamma = (c_{66} / c_{44})^{1/2}$ is the effective mass anisotropy parameter 
%scale for a Larkin domain
equal to the transverse sound speed within the boson analogy.  
Here $c_{44} = n_{\rm B} \tilde\varepsilon_l$ is the tilt modulus 
due to isolated flux lines.
After substitution into Eqs. (\ref{pi}) for the AC response,
the identity (\ref{identity}) 
%in terms of wavenumber/Matsubara frequency
then yields a divergent tilt modulus for the
Bose glass at long wavelength
%/low frequency
%
\begin{equation}
C_{44} (q_z) = n_{\rm B} \tilde\varepsilon_l [1 + (q_z L_*)^{-2}],
\label{C_44}
\end{equation}
with a   longitudinal scale $L_* = \omega_0^{-1}$  that is related to the
transverse Larkin length by the anisotropic  scale transformation $L_* \sim R_c (|) / \gamma$.

Expression (\ref{C_44}) for the uniform  tilt modulus 
of a  Bose glass
is the central result of the paper.
We shall first extract
the Labusch parameter\cite{koopmann}
from the singular behavior that it shows at long wavelength.
In particular,
observe that the elastic energy density
for a periodic tilt of the Bose glass by a displacement
$u_0$ at long wavelength,
${1\over 2} C_{44} (q_z) q_z^2 u_0^2$, 
% $q_z\rightarrow 0$, 
% $2\pi / q_z\rightarrow\infty$
acquires a contribution of the form ${1\over 2} k_0 u_0^2$ from this divergence, 
with  $k_0 = c_{66} / R_c (|)^2$.  The latter is simply the spring constant per unit volume
of the restoring force for a
rigid translation of the Bose glass state about mechanical equilibrium.
Using the estimate
$c_{66} = {1\over 4} \varepsilon_0 n_{\rm B}$
for the shear modulus of the vortex lattice\cite{brandt77}
yields an effective spring constant per vortex line
due to 2D collective pinning limited by plastic creep,
$k_0/n_{\rm B}$,
that is given by
$k_{\rm p} = {1\over 4} \varepsilon_0 / R_c(|)^2$.
Notice that $k_{\rm p}$ depends only weakly on magnetic field.
The Labusch parameter
extracted from the microwave surface resistance
on PLD-YBCO films with nano-rod inclusions also 
exhibits only a weak dependence on external magnetic field
aligned parallel to the nano-rods (or c-axis)\cite{pompeo}!
The above should be compared to the corresponding Labusch parameter
due to point pins\cite{koopmann}, 
which depends strongly on magnetic field.
Indeed, given the conjecture
$k_{\rm p} = (c_{66} / R_c^2 + c_{44} / L_c^2)/n_{\rm B}$
for the Labusch parameter due to 3D collective-pinning
implies that it decays with increasing magnetic field instead as $1/B^2$
in such case.
Here we have assumed anisotropic scaling, and we have used\cite{Tink} $R_c \propto B$.

%Expression (\ref{C_44}) for the uniform  tilt modulus 
%of a  Bose glass
%is the central result of the paper.
We shall next use
Eq. (\ref{C_44}) for the uniform tilt modulus
to test how robust the hexatic Bose glass is to the addition of point pinning centers.
The hexatic Bose glass shows long-range orientational order
in {\it all} directions (see fig. \ref{hex-glass}).
The addition of point pins will therefore break it up into Larkin domains,
of dimensions $R_c^{\prime} \times R_c^{\prime} \times L_c$ 
transverse and parallel to the magnetic induction,
that tilt in the transverse direction
% to the magnetic field
by a distance of order the size  of the vortex core, $\xi$.
%If we now treat the ``pristine'' Bose glass as an elastic medium
%and follow Larkin and Ovchinnikov\cite{LO},
Such a break-up then has an elastic energy cost per unit volume and
a pinning energy gain per unit volume that sum to\cite{LO}\cite{Tink}
\begin{equation}
\delta u = {1\over 2} C_{66} \Biggl({\xi\over{R_c^{\prime}}}\Biggr)^2 
    + {1\over 2} C_{44} \Biggl({\xi\over{L_c}}\Biggr)^2
     - \Biggl({n_0^{\prime}\over{R_c^{\prime 2} L_c}}\Biggr)^{1/2} f_0^{\prime} \xi.
\label{energy}
\end{equation}
Here $C_{66}$ denotes the shear modulus of the hexatic Bose glass, 
which can be approximated by $c_{66}$ in the limit of weak correlated pinning,
while
the corresponding tilt modulus $C_{44}$ is given by expression (\ref{C_44})
evaluated at wavenumber $q_z = 1 / L_c$.
Also, $n_0^{\prime}$ denotes the density of point pins,
while $f_0^{\prime}$ denotes the magnitude of their characteristic force.
Minimizing $\delta u$ with respect to the dimensions of the Larkin domains
then yields standard results for these\cite{Tink}:
%$L_c = 2 C_{44}(q_z) c_{66} \xi^2 / n_0^{\prime} f_0^{\prime 2}$
$L_c = L_c(\cdot) [1+(q_z L_*)^{-2}]$
and
%$R_c = 2^{1/2} [C_{44}(q_z)]^{1/2} c_{66}^{3/2} \xi^2 / n_0^{\prime} f_0^{\prime 2}$.
$R_c^{\prime} = R_c(\cdot) [1+(q_z L_*)^{-2}]^{1/2}$,
where 
$L_c (\cdot) = 2 c_{44} c_{66} \xi^2 / n_0^{\prime} f_0^{\prime 2}$
and
$R_c(\cdot) =  2^{1/2} c_{44}^{1/2} c_{66}^{3/2} \xi^2 / n_0^{\prime} f_0^{\prime 2}$
are respectively the longitudinal Larkin scale and the transverse Larkin scale
in the absence of correlated pinning centers.
The first equation is quadratic in terms of the variable $L_c^{-1}$,
and it has a formal solution
%
%\begin{equation}
$L_c^{-1} = (2 L_c(\cdot))^{-1} + [(2 L_c(\cdot))^{-2} - L_*^{-2}]^{1/2}$.
%\label{L_c}
%\end{equation}
%
The hexatic Bose glass is therefore robust to the addition of weak point pins.
In particular,  
the longitudinal Larkin scale 
$L_c$ remains divergent for $L_c (\cdot) >  L_* / 2$,
which is
equivalent to the inequality
$2^{3/2} R_c (\cdot) >  R_c(|)$.
% to within a numerical factor of order unity.  
%Equation (\ref{L_c}) also indicates that
%the hexatic Bose glass experiences a first-order transition
%to a state that exhibits 3D collective pinning 
%at stronger point pinning
%$2^{3/2} R_c (\cdot) \leq R_c(|)$,
%on the other hand,
%where $L_c \leq L_*$ and $R_c^{\prime} \leq R_c(|)/2$.
%The previous solution (\ref{L_c}) for the longitudinal Larkin scale 
%indicates that correlated pins remain effective
%in the 3D collective-pinning state for $L_c\sim L_*$,
%which is twice as big as $L_c(\cdot)$.
%This regime in 3D collective pinning shall be called a {\it broken} Bose glass.

\section{2D-3D crossover in critical current}
The critical current density of the above hexatic Bose glass, 
which is robust to the addition of  point pins, 
can be obtained
by applying 2D collective pinning theory\cite{jpr-maley}.
All vortex lines can be considered to be rigid rods.
%, including the interstitial ones that are free from material line defects,
%but that are pinned instead by material point defects.
Balancing the Lorentz force against the collective pinning force
over a  Larkin domain\cite{LO}\cite{Tink} 
yields the identity
$J_c  B / c  =    [(n_{\rm p} f_{\rm p}^2 + n_{\rm p}^{\prime} f_{\rm p}^{\prime 2}) / R_c^{2}]^{1/2}$
for the product of   the critical current density along the film, $J_c$,
with the perpendicular magnetic field, $B$,
aligned parallel to the material line defects\cite{jpr-maley}. 
Here $n_{\rm p}$ and $n_{\rm p}^{\prime}$ denote the density of vortex lines pinned by material line defects
and the density of interstitial vortex lines pinned by material point defects,
while $f_{\rm p}$ and $f_{\rm p}^{\prime}$ are the maximum force exerted  on the respective  vortex lines
per unit length.
Again, it is important to observe that the critical current is limited by
{\it plastic creep}
of the vortex lattice due to slip of the quenched-in
lines of  edge dislocations
along their respective glide planes\cite{book}.
Minimization of the sum of the elastic and pinning energy densities
then yields  a higher
% more general expression for the 
density of Larkin domains in the  hexatic Bose glass
with material point defects added by comparison (\ref{ratio}):
$R_c^{-2}   =
  C_0^2 (n_{\rm p} f_{\rm p}^2 + n_{\rm p}^{\prime} f_{\rm p}^{\prime 2}) / (c_{66} b)^2$. 
This reflects  the injection of extra lines of edge dislocations 
that relieve shear stress caused by point pins.
Substitution in turn yields a critical current density,
$J_c = j_c + j_c^{\prime}$,
that has a component due to correlated pins set by the identity
\begin{equation}
j_c  B / c  =  C_0 n_{\rm p} f^2_{\rm p} / c_{66} b,
\label{balance2}
\end{equation}
and that has  a component due to point pins set by the ratio
$j_c^{\prime} / j_c = n_{\rm p}^{\prime} f_{\rm p}^{\prime 2} / n_{\rm p} f_{\rm p}^2$.
The critical current density  notably  varies as
$J_c\propto B^{-1/2}$ with magnetic field in the limit of weak pinning\cite{jpr-maley}.

\begin{figure}
\includegraphics[scale=0.6, angle=-90]{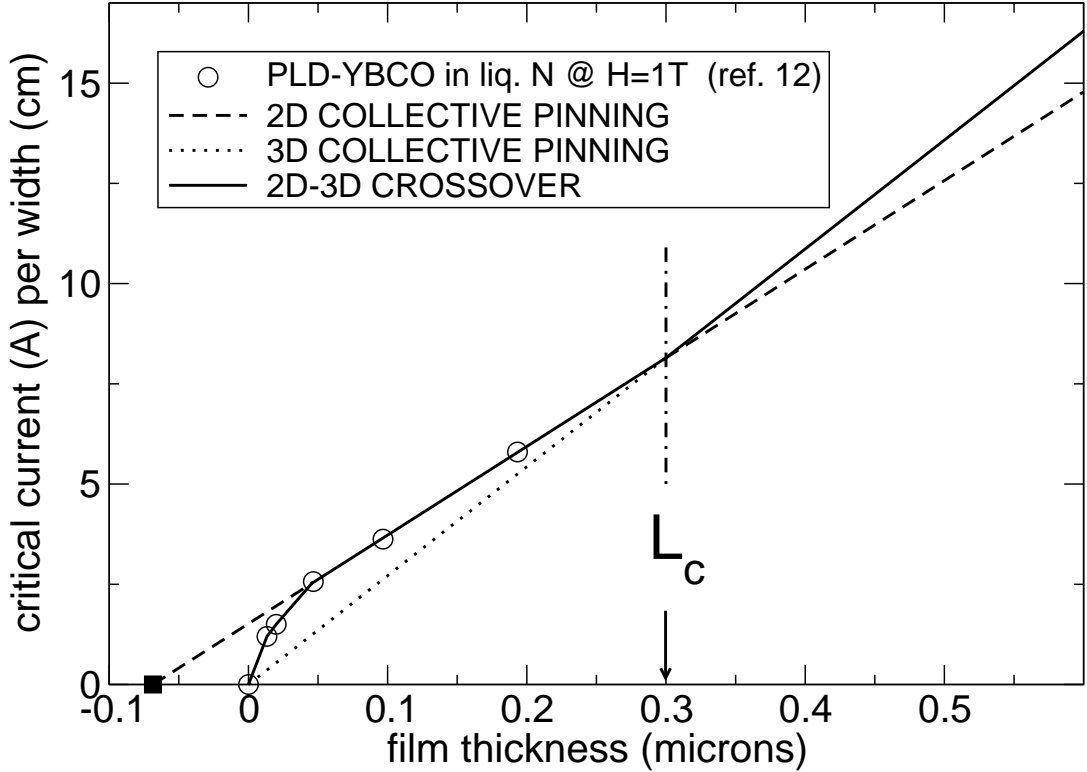}
\caption{Shown is the dependence on film thickness predicted for the critical current of
a defective vortex lattice found in a broken Bose glass state (solid line).
The dashed and dotted lines are extrapolated from the 2D and 3D behaviors, respectively.  
%Data taken from  
Measurements 
made by Sang Kim (circles, ref. \cite{kim-larbalestier})
of the critical current
on a thin film of PLD-YBCO
at liquid nitrogen temperature  
subject to $1\, {\rm T}$ magnetic field
aligned parallel to the c-axis
are fit to the straight dashed line predicted by 2D collective pinning (ref. \cite{jpr-maley}).
This yields an  intercept $-\tau_0 = -69\, {\rm nm}$ and a slope $j_c = 0.22\, {\rm MA/cm}^2$.
Although the value of $L_c$ shown here is indeed larger than the lower bound $L_* = 24\, {\rm nm}$ 
[see Eq. (\ref{L*})]
and is consistent with the fit to 2D collective pinning, it is only hypothetical.}
\label{2d-3d}
\end{figure}

Consider now a film geometry of thickness $\tau$ along the  axis of
the material line defects.
%the magnetic field.
The forces due to point pins 
add up statistically along a   rigid interstitial vortex line.
The effective pinning force per unit length experienced by an interstitial
vortex line  is then given by\cite{kes-tsuei}
$f_{\rm p}^{\prime} =  
f_0^{\prime} /(\tau_{\rm p}^{\prime} \tau )^{1/2}$
at film thicknesses $\tau$
%$\tau > \tau_{\rm p}^{\prime}$if the film is 
that are much greater than 
the average separation $\tau_{\rm p}^{\prime}$ between such pins along the field direction. 
Again, $f_0^{\prime}$ denotes the maximum force exerted by a point pin.
The relative 
contribution by point pins to the critical current density
is then predicted to show an inverse dependence 
on film thickness,
$j_c^{\prime} / j_c = \tau_0 /\tau$,
that is set by the scale
$\tau_0 = 
(n_{\rm p}^{\prime}/n_{\rm p})(f_0^{\prime 2}/f_{\rm p}^2 \tau_{\rm p}^{\prime})$.
%%4\pi r_0^{\prime 2} s^2 n_{0}^{\prime}$.
%Here  $s$ denotes the separation between adjacent layers in YBCO, 
%and $f_0 = f_{\rm p} s$.
We thereby obtain the linear dependence on film thickness $I_c^{(2D)} = (\tau_0 + \tau) j_c$
for the net critical current per unit width,
$\tau J_c$. 
Last, comparison of Eq. (\ref{ratio}) with Eq. (\ref{balance2}) yields the useful expression
\begin{equation}
L_* = \gamma^{-1} [(3^{5/4}/2^{7/2} C_0)(\xi\cdot a_{\rm vx}) (j_0 / j_c)]^{1/2}
\label{L*}
\end{equation}
for the longitudinal scale characteristic of the Bose glass 
as a function of the bulk critical current density, $j_c$. 
Here  $a_{\rm vx} = n_{\rm B}^{-1/2}$ is the average distance between vortex lines
and $j_0$ is the depairing current density (see ref. \cite{Tink}).
Figure \ref{2d-3d} shows a fit to data for the critical current versus thickness
obtained from
a thin film of PLD-YBCO at liquid-nitrogen temperature
in $1\, {\rm T}$ magnetic field 
aligned parallel to the $c$ axis\cite{kim-larbalestier}.
A bulk critical current density $j_c = 0.22\, {\rm MA/cm}^2$ is extracted from it.
Using a value of
$j_0 = 36\,{\rm MA/cm}^2$
for the depairing current (ref. \cite{kim-larbalestier}), 
of $\xi = 11\,{\rm nm}$ for the coherence length, of $\gamma = 7$ for the mass
anisotropy parameter, and setting $C_0=1$ 
yields a longitudinal scale $L_* = 24\,{\rm nm}$.

Finally, consider again
an  arrangement of material line defects 
that are aligned along the $c$ axis,
that show no voids or clusters in the transverse directions,
but that are  broken up into relatively long rods  of length $L_0\gg L_*$.
It can be realized by meandering material line defects\cite{dam}\cite{klaassen}, 
where $L_0$ is the correlation length for alignment along the $c$-axis, 
or by artificial nano-rod defects\cite{bzo}\cite{goyal}.
Expression (\ref{C_44}) for the divergent tilt modulus indicates
proximity to the regime of 2D collective pinning, where both $L_c$ and $L_0$ are divergent.
It hence indicates that $L_c$ is  also large compared to $L_*$ here.  
Second, observe that $L_* \sim R_c(|)/\gamma$ coincides with 
the longitudinal Larkin scale if the  rods are considered to be point defects.  
The previous ultimately implies  the chain of inequalities $L_0 \geq L_c \gg L_*$.  
They are consistent with a {\it broken} Bose glass state for the vortex lattice,
which is threaded by isolated lines of edge dislocations of length $L_c$ along the $c$-axis that are
connected together by lines of 
screw-dislocations\cite{book} along the transverse directions\cite{wordenweber-kes}.
As in the limiting case of the true Bose glass state (fig. \ref{hex-glass}), 
the lines of edge dislocations 
%(parallel to the $c$-axis) 
are injected
into the pristine vortex lattice in order to relieve shear stress due to the correlated pins.
Larkin domains hence are finite volumes\cite{LO}\cite{Tink}.  
In contrast to their transverse dimensions, however,
Larkin domains exhibit well defined boundaries along 
the direction parallel to the correlated pinning centers,
across which the vortex lattice slips by a Burger's vector
due to the presence of the screw dislocations\cite{book}.
The critical current density expected from this peculiar example of 3D collective pinning 
therefore coincides with that of a Bose glass of thickness $L_c$: 
$J_c = j_c + j_c^{\prime}$, with a {\it bulk} component due to interstitial vortex lines
$j_c^{\prime} / j_c = \tau_0 /L_c$.  
The critical current per unit width then varies with film thickness as $I_c^{(3D)} = \tau J_c$,
showing no offset.
Figure \ref{2d-3d} depicts the predicted dependence on film thickness
for the critical current of such a broken Bose glass.
It notably exhibits 2D-3D cross-over at film thicknesses in the vicinity of $L_c$ \cite{wordenweber-kes}.

\section{Concluding remarks}
The dependence of the critical current on thickness $\tau$
shown by certain films of PLD-YBCO 
is in fact consistent with dimensional cross-over
at $\tau\sim 1\, \mu {\rm m}$. 
A previous attempt 
by Gurevich
to account for such behavior 
by collective pinning of individual vortex lines at point defects 
%in the single-vortex regime (one pinned vortex line per Larkin volume) 
yields a longitudinal Larkin scale $L_c\sim 10 \,{\rm nm}$ that is too small, 
however\cite{kim-larbalestier}.
We find here, on the other hand, that pinning due to 
correlated material defects of length $L_0$
%still exhibits 3D collective pinning,
yields a  Larkin scale $L_c$ that is much longer than 
that [$L_* = 24\,{\rm nm}$, see Eq. (\ref{L*})] 
expected from point pins of comparable strength
if $L_0$ is much longer than that scale as well.  
Indeed, we predict here that the film thickness at which the critical current crosses over
from 2D to 3D behavior is of order the effective length of the correlated pinning centers
when that length satisfies $L_0 \gg L_*$.

%In conclusion, 
%although a rigid vortex lattice that experiences correlated material defects of finite length $L_0$
%still exhibits 3D collective pinning,
%its longitudinal Larkin scale is long compared to
%that ($L_*$) expected from point pins of comparable strength
%if $L_0$ is long compared to that scale as well.  
%%The micro-structure just described
%%resembles  that found in films of PLD-YBCO\cite{huij}\cite{bzo}\cite{goyal}.  
%%The broken Bose glass decribed above
%%is then a good candidate for the state of the vortex lattice
%%that exists in such type-II superconductors at applied magnetic field.

\acknowledgments  The author thanks 
Leonardo Civale, Chandan Das Gupta and Sang-il Kim for discussions.
This work was supported in part by the US Air Force
Office of Scientific Research under grant no. FA9550-06-1-0479.

\end{document}